\documentclass[a4paper,10pt]{article}
\usepackage{amsmath,amsthm,latexsym,amssymb,amsfonts}

\newcommand{\R}{{\mathbb R}}
\newcommand{\C}{{\mathbb C}}

\newcommand{\hgr}{{\cal H}_{\rm gr}}
\newcommand{\hsc}{{\cal H}_{\rm sc}}
\newcommand{\hkin}{{\cal H}_{\rm kin}}
\newcommand{\hphys}{{{\cal H}_{\rm phys}}}
\newcommand{\hilb}{{\mathcal H}}

\begin{document}

\title{The status of Quantum Geometry in the dynamical sector
of Loop Quantum Cosmology}
\author{ Wojciech
Kami\'nski$^{1}$\thanks{wkaminsk@fuw.edu.pl},
        Jerzy Lewandowski$^{1,2}$\thanks{lewand@fuw.edu.pl} and
        {\L}ukasz Szulc$^{1}$\thanks{lszulc@fuw.edu.pl}}
\date{\it 1. Instytut Fizyki Teoretycznej,
Uniwersytet Warszawski, ul. Ho\.{z}a 69, 00-681 Warszawa, Poland\\
2. Physics Department, 104 Davey, Penn State, University Park, PA 1602, USA \\
[.5cm]} \maketitle
\begin{abstract}
This letter is motivated by the recent papers by Dittrich and  Thiemann and,
 respectively, by Rovelli discussing  the status of Quantum Geometry in
 the dynamical sector of Loop Quantum Gravity. Since the papers consider
 model examples, we also study the issue in the case of an example, namely
 on the Loop Quantum Cosmology model of space-isotropic universe. We derive
 the Rovelli-Thiemann-Ditrich partial observables corresponding to the
 quantum geometry  operators of LQC in  both Hilbert spaces: the
 kinematical one and, respectively, the physical Hilbert space of solutions to the quantum
 constraints. We find, that Quantum Geometry can be used to characterize the
 physical solutions, and the operators of quantum geometry preserve
 many of their kinematical properties.
 \end{abstract}

\section{Introduction} One of the issues of the canonical gravity is the
lack of  explicit formulae for the Dirac observables. In the consequence,
a role the kinematic quantization of the gravitational field
plays after implementing the quantum Einstein constraints is not
known. In  Loop Quantum Gravity \cite{LQGrevsbooks}
 the operators representing the intrinsic
3-geometry of a given Cauchy surface as well as the extrinsic
curvature are known \cite{QG}. Their properties, spectra, eigenvalues
and eigenfunctions were studied \cite{Qgprop}. But what is their meaning in
the dynamical theory? The bottom line is that the kinematical   operators
are used to define the quantum constraint operators \cite{QSD}. Therefore
their relevance is unquestionable. However, the open question is which
properties of the kinematical geometry operators and other structures of the
kinematical quantum theory are preserved by the passage to the Dirac
observables. New insights were given recently by a work by
Ditrich and Thiemann \cite{DTnewest}. They study various toy
examples of the explicit construction of the Dirac observables by
using the so called "partial observables" method \cite{ParObs}. That
method allows one to construct a Dirac observable from any
kinematical observable. It is shown in \cite{DTnewest}, however,
that the discreteness of the kinematical operators does not imply
the discreteness of the corresponding quantum Dirac observables and
vice versa. The procedure of turning a kinematical observable into
the dynamical one can wash out all the properties and replace them
by others. A few days after the Ditrich and Thiemann's paper
appeared in the archives, Rovelli send his response \cite{Rovnewest}.
According to Rovelli,  the  examples of \cite{DTnewest}
are too distant from the Loop Quantum Gravity.
\medskip

Those recent works motivated us to check the status of the issue of
the quantum Dirac observables in the model of LQG called Loop
Quantum Cosmology \cite{Abhaynewest,LQC}. We consider
in this work on the best understood, "improved"  LQC model constructed
from the family of the space-isotropic gravitational fields
(the Friedman-Robertson-Walker spacetimes) coupled to a space-isotropic
massless scalar fields
\cite{APS}. The model has two advantages: on the one hand, it
has a lot of the properties of  LQG, and is understood as a toy model
of LQG. Therefore, it is hopped that many results concerning LQC should
admit generalizations to LQG. On the other hand, the
model is simple enough to be be quite well understood. In particular,
the quantum observables of this model can be derived explicitly.
This is what we do in the paper. The specific question we focus on is the
role of the quantum geometry of LQC in the space of the solutions to
that theory. A discussion of the technical subtleties
related to our form of the scalar constraint which was adapted here to the
question we are studying is contained in the last section.

\bigskip

\section{The APS model, positive frequencies}
The kinematical Hilbert space $\hgr$ of the gravitational degrees of
freedom in the FRW-LQC model is spanned by the basis of orthonormal
vectors $|v)$, labeled by all the possible real values of $v\in\R$,
\begin{equation}
(v|v')\ =\ \delta_{v,v'}\ =\ \begin{cases}1,&  {\rm if}\ \
v=v'\\0,&\ \ {\rm otherwise}\end{cases}
\end{equation}
That is, the Hilbert space and the scalar product
$(\cdot|\cdot)_{\rm gr}$ are
\begin{align}
\hgr\ &=\ \{ \sum_{i=1}^\infty a_i|v_i) :\  a_i\in \C, \ \
\sum_{i=1}^\infty |a_i|^2<\infty \}\nonumber\\
(\sum_{i=1}^\infty a_i|v_i)\ |\ \sum_{j=1}^\infty b_j|v_j))_{\rm
gr}\ &= \sum_{i=1}^\infty
\sum_{j=1}^\infty\overline{a_i}b_j\delta_{v_i,v_j}
\end{align}

The kinematical observables are the quantum volume operator
\begin{equation}
\hat{V}|v\rangle\ =\ v|v\rangle
\end{equation}
and the "improved" \cite{APS} quantum holonomy operator
\begin{equation}
\hat{h}_\lambda |v\rangle\ =\ |v+\lambda\rangle, \ \ \ \lambda\in\R.
\end{equation}

The volume operator  $\hat{V}$ represents the 3-volume of an
isotropic space like section of the universe in the closed FRW case,
or some fixed box in an isotropic space like section of the universe
the open FRW case. The quantum holonomy operator represents a
kinematical observable involving the extrinsic curvature of an
isotropic section (we skip some constants and details that can be
found in \cite{APS}).

 The kinematical Hilbert
space of the scalar field is the space of the square integrable
functions on $\R$ endowed with the Lebesgue measure,
\begin{equation}
\hsc\ =\ {\rm L}^2(\R).
\end{equation}
The scalar field operator is just the multiplication,
\begin{equation}
(\hat{\Phi}\psi)(\phi)\ =\ \phi\psi(\phi).
\end{equation}
The scalar field momentum operator $\hat{\Pi}$ is
\begin{equation}
\hat{\Pi}\psi\ =\ \frac{1}{i}\frac{d}{d\phi}.
\end{equation}

Finally, the kinematical Hilbert space of the isotropic
gravitational field coupled to the isotropic scalar field is
\begin{equation}\label{hkin}
\hkin\ =\ \hsc\otimes\hgr.
\end{equation}
And the kinematical observables are the following operators
\begin{equation}
1\otimes\hat{V},\ \  1\otimes\hat{h}_{\lambda},\ \ \hat{\Phi}\otimes
1,\ \  \hat{\Pi}\otimes 1,
\end{equation}
where $\lambda\in\R$ runs through all the set $\R$. There are only 2
degrees of freedom, hence the declared set of the
`momentum' observables $\hat{h}_{\lambda}$ is overcomplete. The
reason is, that the holonomy operators are unitary. If instead of
the operator $\hat{\Pi}$ we were using an operator $e^{-i\alpha
\hat{\Pi}}$ we would also admit all the values of $\alpha$.
\bigskip

We turn now to the dynamics.  The dynamics of the theory is given by
the scalar constraint operator $\hat{C}$. The massless scalar field
we consider here  is the best understood case. Let us start with the
simplest from the point of view of our work formulation of the
constraint. The scalar constraint used in \cite{APS} can be written
in the following form
\begin{equation}
\hat{C}_-\ =\ \hat{\Pi}\otimes 1 - 1\otimes H,
\end{equation}
where $H$ is an operator defined in the kinematical Hilbert space
$\hgr$ of the gravitational degree of freedom, it does not involve
any of the operators acting in $\hsc$. The operator $H$ is not
diagonal, that is it does not commute with $\hat{V}$,
$$ [\hat{V},H]\ \not=\ 0. $$

For the sake of completeness let us consider here the case of a
constraint operator of the following form
\begin{equation}\label{Cpm}
\hat{C}_{\pm}\ =\ \hat{\Pi}\otimes 1 \pm 1\otimes H,
\end{equation}
where we fix either $+$ or $-$.

A strong Dirac observable is an operator  commuting with the
constraint $\hat{C}$ in (a sufficiently large domain in) $\hkin$.
Certainly
\begin{equation}
[\hat{\Pi},\hat{C}_\pm]\ =\ 0,
\end{equation}
hence the scalar field quantum momentum is a Dirac observable.

We are particularly interested in  those observables which involve
the operators acting in the kinematical Hilbert space of the
gravitational degrees of freedom which define the quantum geometry:
the quantum volume operator $\hat{V}$ and the  quantum holonomy
operators $\hat{h}_\lambda$. None of them is an observable,
$$ [\hat{V},\hat{C}_\pm]\ \not=\ 0\ \not=\ [\hat{h}_\lambda,\hat{C}_\pm].$$
This is a model version of the outstanding problem in LQG: the
quantum geometry has been defined on the kinematical level. The
operators of the quantum geometry do not commute with the
constraint. {\it However}, in the case of either of the constraint
$\hat{C}_-$ or $\hat{C_+}$,  it is easy to assign a Dirac observable
${\cal O}_{\pm}$ to any given operator ${\cal O}$  in $\hgr$.
Indeed, fix any number $\phi_0\in\R$ of the operator scalar field
operator $\hat{\Phi}$, and define
\begin{equation}\label{Vpm}
{\cal O}_{\pm,\phi_0}\ :=\ e^{\pm i{(\hat{\Phi}-\phi_0)\otimes H}}
1\otimes {\cal O} e^{\mp i{(\hat{\Phi}-\phi_0)\otimes H}}
\end{equation}
(we could just fix $\phi_0=0$ but each choice of $\phi_0$ will have
a natural  interpretation). It is easy to check, that the result is
an operator in $\hkin$ which satisfies
\begin{equation}
[{\cal O}_{\pm,\phi_0},\hat{C}_\pm]\ =\ i\,e^{\pm
i(\hat{\Phi}-\phi_0)\otimes H} \left( \pm 1\otimes\left[{\cal
O}_{\pm,\phi_0},H\right]\ \mp \ 1\otimes\left[{\cal
O}_{\pm,\phi_0},H\right]\right) e^{\mp i(\hat{\Phi}-\phi_0)\otimes
H}\ =\ 0.
\end{equation}
That form (\ref{Vpm}) of a Dirac observable is not a surprise,
because the APS constraint $\hat{C}_\pm$ has exactly the same form
as the Rovelli-Schroedinger constraint \cite{Rovnewest,RovQM} with the
Rovelli time operator $\hat{t}$ replaced with the scalar field
operator $\hat{\Phi}$.

The Hilbert space  $\hphys_{\pm}$ of "solutions" to the quantum
constraint defined by one of the operators  $\hat{C}_{\pm}$
(\ref{Cpm}) is identified \cite{APS} with the space of
$\hgr$--valued functions defined on the spectrum $\R$ of the scalar
field operator,
\begin{equation}\label{solpm}
\R\ni\phi\mapsto \psi(\phi)\in\hgr,
\end{equation}
which satisfy the equation
\begin{equation}\label{Schr}
-i\frac{\partial}{\partial \phi}\psi\ =\ \pm H\psi.
\end{equation}
The scalar product between two solutions is defined by the scalar
product in $\hgr$ calculated at any value of $\phi$ due to the
identity
\begin{equation}
(\psi(\phi)\ |\ \psi'(\phi))_{\rm gr}\ =\ (\psi(\phi')\ |\
\psi'(\phi'))_{\rm gr}.
\end{equation}

Now, it is easy to see,  an general operator acting in $\hgr$ does
not admit a unique action on a solution (\ref{solpm}), unless it commutes
with $H$. The action can be defined only at fixed value of
$\phi-\phi_0$:  given a solution (\ref{solpm}) and an operator
${\cal O}$, consider another solution
\begin{equation}
\R\ni\phi\mapsto \psi"(\phi)\in\hgr,
\end{equation}
such that
\begin{equation}\label{O1}
\psi"(\phi_0)\ =\ {\cal O}\psi"(\phi_0).
\end{equation}
 On the other hand, the action of the corresponding  Dirac observable operator
 ${\cal O}_{\pm,\phi_0}$ is well defined on each solution (see below why), and it
 is
\begin{equation}\label{O2}
{\cal O}_\pm\psi(\phi)\ =\ e^{\pm i(\phi-\phi_0) H}{\cal O}e^{\mp
i(\phi-\phi_0) H}\psi(\phi).
\end{equation}
The two actions (\ref{O1}, \ref{O2}) coincide.

The operator $H$ commutes with itself, and its action unambiguously
passes to the space of solutions (\ref{solpm}). In the space of
solutions, the action of the scalar momentum operator $\hat{\Pi}$
becomes
$$\hat{\Pi}\ =\ \pm H.$$
\medskip

At this point we are in the position to address the main issue of
the paper, the issue of the role of the status of the kinematical
Hilbert space of the gravitational degrees of freedom and quantum
geometry for the physical space of solutions to the quantum
constraints. We observe that:
\begin{itemize}
\item The kinematical Hilbert space  of the gravitational degrees of
freedom  is unitarily equivalent to  the space of the physical
solutions.
\item Every Dirac observable $\hat{\cal O}_{\pm,\phi_0}$ constructed from an operator
${\cal O}$ in $\hgr$ is mathematically the same as  the Heisenberg
picture of the operator ${\cal O}$ defined by the equation
(\ref{Schr}) understood as the Schroedinger equation. Hence it is
unitarily equivalent to the original operator ${\cal O}$.
\item Every quantum geometry operator (defined in $\hgr$)  itself can be
used in the physical Hilbert space and its status is exactly the
same as the the status of any  operator in the Schroedinger quantum
mechanics. In particular the quantum volume operator, despite of
non-commuting with the constraint, has the same status in the space
of the physical solutions as the position operator in Quantum
Mechanics of a point particle whose dynamics is governed by the
quantum Hamiltonian operator $H$.
\item Eventually, in this case we would not need to invoke the Dirac
observables theory at all, just use the QM  framework!
\end{itemize}
 
\section{An equivalent, systematic construction} The construction of the space
${\hphys}_\pm$ of the solutions (\ref{solpm},\ref{Schr}) to the
constraint (\ref{Cpm}) can be performed in a systematic by using the
scheme proposed by Thiemann in the context of the master constraint
operator \cite{Thiemmaster}. One begins with the spectral
decomposition of the kinematical Hilbert space defined by the
spectral decompositions of the Hilbert spaces $\hsc$ and $\hgr$
corresponding to the operators $\hat{\Pi}$ and $H$ respectively. That
is, an element of $\hkin$ is identified with an assignment
\begin{equation}\label{specdecel}
\R\times\R \ni (\pi,E)\ \mapsto\ \psi(\pi,E)\in
\hilb_\pi^{\hat{\Pi}}\otimes\hilb_E^{H},
\end{equation}
where $\hilb_\pi^{\hat{\Pi}}$ and $\hilb_E^{H}$ are some Hilbert
spaces assigned to the numbers $\pi$ and $E$, respectively.  The
kinematical scalar product in  $\hkin$ reads
$$ (\psi|\psi')_{\rm kin}\ =\ \int d\pi dE
({\psi(\pi,E)}|\psi'(\pi,E))_{\pi,E},$$
where $d\pi$ and $dE$ are some measures and $(\cdot|\cdot)_{\pi,E}$
is the  scalar product in $\hilb_\pi^{\hat{\Pi}}\otimes\hilb_E^{H}$.
Finally, the action of the operators $\hat{\Pi}$ and, respectively
$H$ in this representation reads
\begin{equation}
(\hat{\Pi}\psi)(\pi,E)\ =\ \pi\psi(\pi,E), \ \ \ (H\psi)(\pi,E)\ =\
E\psi(\pi,E).
\end{equation}
In fact
$$\hilb_\pi^{\hat{\Pi}}\ =\ \C,$$
and $d\pi$  is the Lebesgue measure. The measure $dE$ is also known
in a large class of cases  \cite{KL}.

In this formulation, a solution to the constraint  (\ref{Cpm}) is
just the restriction of the definition (\ref{specdecel}) to the
subset of $\R\times\R$ such that
$$\pi\ =\ \mp E,$$
(either $+$ or $-$, depending on the sign in (\ref{Cpm})), that is a
solution is a map
\begin{equation}\label{sol}
\R \ni E\ \mapsto\ \psi(\mp E,E)\in \hilb_{\mp
E}^{\hat{\Pi}}\otimes\hilb_E^{H}.
\end{equation}
The solutions set a vector space. The  scalar product between two
solutions is defined to be
$$ (\psi|\psi')_{\rm phys}\ =\ \int dE
({\psi(\mp E,E)}|\psi'(\mp E,E))_{\mp E,E}.$$
  The solutions and the scalar
product define the physical Hilbert space ${\hphys}_{\pm}$. We would
like to induce in ${\hphys}_{\pm}$ an action of some of the
operators introduced in $\hkin$. To do so, we identify  every
solution (\ref{sol}) with a linear functional $\langle\psi|$
(defined in some domain $D_{\pm}$ in $\hkin$ -- we do not bother the
reader with the domains of the operators in this paper, but we have
to mention this at that point) by
\begin{equation}
D_{\pm}\ni \psi' \mapsto \langle \psi|\psi'\rangle\ =\ \int dE
(\psi(\pm E,E)|\psi'(\mp E,E))_{\mp E,E}.
\end{equation}
Every operator in $\hkin$ whose range contains $D_{\pm}$, by the
duality,  maps each solution (\ref{sol}) into another linear
functional in $\hkin$. If this map preserves the form (\ref{sol}),
then the operator induces an operator in $\hphys$. In particular,
every operator commuting with the constraint $\hat{C}_{\pm}$
preserves the space of solutions (\ref{sol}). Therefore again, the
Dirac observables defined in the previous section pass to
${\hilb}_\pm$.

 This description of ${\hphys}_{\pm}$ and the action
of the Dirac observables is unitarily equivalent to that used in the
previous section.

\section{The APS model, all the frequencies}
Now, we can turn to the full scalar constraint of the LQC-FRW case.
The full scalar constraint operator reads
\begin{equation}
\hat{C}\ =\ \hat{\Pi}^2\otimes 1 - 1\otimes H^2.
\end{equation}
We can write
\begin{equation}\label{C-C+}
\hat{C}\ =\ (\hat{\Pi}\otimes 1 + 1\otimes H)(\hat{\Pi}\otimes 1 -
1\otimes H)\ =\ \hat{C}_{+}\hat{C}_-,
\end{equation}
where the factors commute. One can conjecture, that the space of
solutions of the constraint (\ref{C-C+}) consists of the solutions
to the constraint $\hat{C}_-$ and the solutions to the constraint
$\hat{C}_+$. Indeed, the conjecture is true in the following sense.
Consider again the representation of the elements of $\hkin$ by the
spectral decomposition (\ref{specdecel}). Now, a solution to the
quantum constraint defined by the constraint operator (\ref{C-C+})
corresponds to the restriction of the definition (\ref{specdecel})
to the subset
$$ \{(\pi,E)\in\R\times\R\\ :\ \pi^2=E^2\}$$
%
that is an assignment
\begin{equation}\label{sol+-}
\{(\pi,E)\in\R\times\R\\ :\ \pi^2=E^2\} \ni (\pi,E)\ \mapsto\
\psi(\pi,E)\in \hilb_\pi^{\hat{\Pi}}\otimes\hilb_E^{H},
\end{equation}
Certainly, each of the solutions (\ref{solpm}) is a solution in the
sense of (\ref{sol+-}). Moreover, every solution (\ref{sol+-}) is a
sum of a pair of solutions (\ref{solpm}), one  to  the constraint
$\hat{C}_-$, and the other one to $\hat{C}_+$. That decomposition of
(\ref{sol+-}) is unique, orthogonal, and the components are
independent, provided we assume that the subset $\{0\}\subset\R$ is
of measure $0$ according to $dE$. In that case, the physical Hilbert
space is
\begin{equation}\label{decomp}
\hilb_{\rm phys}\ =\ \hilb_{{\rm phys},-}\oplus \hilb_{{\rm
phys},+},
\end{equation}
where $\hilb_{{\rm phys},\pm}$ stands for the Hilbert space of
solutions to the constraint $\hat{C}_\pm$.

In this case of the constraint (\ref{C-C+}), defining explicitly a
large class of (quantum) Dirac observables in $\hkin$ is not so easy
as before. Still the operator $\Pi$ defined in $\hkin$ passes to
$\hphys$ and induces therein the following operator
$$ \Pi_{\rm phys}= H_-\ -\ H_+, $$
where $H_\pm$ annihilates the term ${\hphys}_\mp$  in (\ref{decomp})
whereas
$$ H_\pm |_{{\hphys}_\pm} \ =\ H.$$

Given a quantum geometry operator ${\cal O}$ defined in $\hgr$, a
counterpart of the Dirac observable (\ref{Vpm}) derived along the
Rovelli-Thiemann-Dietrich  method would heuristically look as
\begin{equation}\label{dirheur}{\cal O}_{\phi_0}\ =\
"e^{i(\hat{\Phi}-\phi_0)\hat{\Pi}^{-1}\otimes H^2}1\otimes{\cal O}
e^{-i(\hat{\Phi}-\phi_0)\hat{\Pi}^{-1}\otimes H^2}".
\end{equation}
But completing this definition is not easy.

However, in the previous section we have seen a more general
condition on an operator defined in $\hkin$ which ensures that the
operator naturally induces an operator in $\hphys$. To formulate it
in the current case, we turn each solution (\ref{sol+-}) into a
linear functional $\langle\psi|:D\rightarrow\C$  (defined in some
domain $D\subset\hkin$), such that
\begin{equation}
\langle \psi|\psi'\rangle\ =\ \int dE (\psi( E,E)|\psi'( E,E))_{
E,E}\  + \int dE(\psi(-E,E)|\psi'(-E,E))_{-E,E}.
\end{equation}
An operator  in $\hkin$ whose domain contains $D$ maps acts by the
duality,
$$ \langle{\cal O}\psi|\ =\ \langle\psi|{\cal O}. $$
The condition is, that the right hand side be again of the solution
form (\ref{sol+-}).

Those who enjoy exploring heuristic formulae can proceed as follows.
Assume for this paragraph, that it makes sense to make the following
replacement  in (\ref{dirheur})
$$\hat{\Pi}^{-1}\ \ {\rm replaced\ \ by}\ \ \pm H^{-1}. $$
Then, the heuristic formula "restricted" to (\ref{sol+-}) becomes
\begin{equation}\label{heur2}{\cal O}_{\phi_0}|_{\hphys_-\oplus\hphys_+}\ =\
e^{\pm i{(\hat{\Phi}-\phi_0)\otimes H}} 1\otimes {\cal O} e^{\mp
i{(\hat{\Phi}-\phi_0)\otimes H}},
\end{equation}
where the upper/lower sign corresponds to $\hphys_\mp$.

The exact form of that  conclusion, is that given an operator
$1\otimes {\cal O}$ in $\hkin=\hsc\otimes\hgr$,  the corresponding
Dirac observable operator ${\cal O}_{\phi_0}$ can be defined in
$\hkin$ by the assumption, that in the spectral decomposition
(\ref{specdecel}) representation, in a neighborhood of the lines
$$\pi = E, \ \ {\rm or}\ \ \pi=-E$$
it equals (\ref{heur2}), and otherwise it is arbitrary.

The resulting operator in $\hphys=\hphys_-\oplus\hphys_+$ has the
expected form,
$${\cal O}_{\phi_0}\ =\ {\cal O}_{\phi_0,-} + {\cal O}_{\phi_0,+},$$

\section{Conclusions}  At the end of the first section we have already
itemized  our conclusions concerning  the  eventual role of the
kinematic quantum geometry operators in the case of the first example.
The "physical" Hilbert space ${\hphys}_+$ is the space of the positive  frequency
solutions of the APS model quantum scalar constraint (appropriately adapted,
see below). Each quantum solution can be thought of as an evolving state
of the (kinematical) quantum geometry. In the consequence, the physical
Hilbert space ${\hphys}_{+}$ is unitary with the kinematical Hilbert space
$\hgr$ of the quantum excitations of the space time geometry, that is
$${\hphys}_+ \ \cong\  \hgr. $$
However, an isometry
$$ U_{\phi_0}: {\hphys}_+ \rightarrow  \hgr $$
is not unique. It depends on a value  $\phi_0$ of
the scalar field. One can interpret that dependence as "evolution" and
think of the parameter $\phi_0$ as  time emerging from LQC.
In this case, $\phi_0$ can be identified with an element of the spectrum of
of the kinematical scalar field operator, however it is not clear how
general is that observation. In a consequence, an extension to $\hphys_+$ of an
operator ${\cal O}$ of quantum geometry defined in $\hgr$ is well defined
at every "instant" of that evolution. Hence we may denote it by ${\cal
O}_{+,\phi_0}$. The resulting operator, in terms of  the QM
analogy, is the Heisenberg picture of ${\cal O}$. On the other hand,
we calculate the Rovelli-Thiemann-Ditrich quantum observable operator in $\hkin$
assigned to ${\cal O}$ upon the choice of the scalar field $\phi$ as a time, and $\phi_0$ as the
instant of time. The result is, that the RTD observable extended to
${\hphys}_+$ just coincides with ${\cal O}_{\pm,\phi_0}$. From the mathematical
point of view, that interpretation is just equivalent to the Rovelli formulation
of Quantum Mechanics \cite{RovQM}. What is important for us in the current
paper, is that the equivalence applies directly to Quantum Geometry in
this  LQC model.

We also consider the version of the APS model which admits both, positive
and negative frequencies. Upon the assumption that there is no
normalizable zero frequency mode (this is true in the $(k=0,1), (\Lambda\le 0)$ cases)
the physical Hilbert space naturally
splits
\begin{equation}\label{decomppm}
\hphys\ =\ {\hphys}_-\oplus{\hphys}_+\ \cong\ \hgr\oplus\hgr,
\end{equation}
and whereas the first isometry is natural, the second again depends
on a value $\phi_0$. The action of each quantum geometry operator
$\cal O$ can be extended to $\hphys$ in the diagonal manner, by
using this identification in every given instant $\phi_0$,
$$ {\cal O}_{\phi_0}\ =\ {\cal O}_{\phi_0,-}\ +\ {\cal
O}_{\phi_0,+}.$$
In this case the constraint operator is no longer linear in the scalar
field momentum operator, hence we were not able to  define
in $\hkin$ explicitly the Rovelli-Thiemann-Dietrich quantum observable
operator assigned to the operator ${\cal O}$. However, we proposed a more
general definition of the quantum observable and applied it to a
heuristic quantization of a classical RDT quantum observable. The result
again coincides with ${\cal O}_{\phi_0}$.

We have postponed until the end of this paper remarks concerning
technical details of the scalar constraint. The original gravitational
scalar constraint operator in the kinematical Hilbert space $\hkin$
(\ref{hkin}) has the following form
\begin{equation}
\hat{C}\ =\ \hat{\Pi}^2 \otimes \widehat{v^{-1}}\ +\
1\otimes\hat{C}_{\rm gr},
\end{equation}
where $\widehat{v^{-1}}$ is a  quantum inverse volume operator. Whereas the
self-adjointness of the operator $\hat{C}$ can be proven quite generally
\cite{KL}, the first term does not commute with the second one. For this
technical reason, it is reasonable to consider instead the operator
\begin{equation}\label{C1}
\sqrt{\widehat{v^{-1}}^{-1}}\circ\hat{C}\circ
\sqrt{\widehat{v^{-1}}^{-1}}\ =\ \hat{\Pi}^2 \otimes 1\ +\
1\otimes\sqrt{\widehat{v^{-1}}^{-1}}
\hat{C}_{\rm gr}\sqrt{\widehat{v^{-1}}^{-1}}.
\end{equation}
On the other hand, in the original APS model another formulation of the
constraint operator is considered, namely
\begin{equation}\label{C2}
\widehat{v^{-1}}^{-1}\circ\hat{C}\ =\ \hat{\Pi}^2 \otimes 1\ +\
1\otimes {\widehat{v^{-1}}^{-1}}
\hat{C}_{\rm gr},
\end{equation}
and the operator is symmetric in a space ${\hkin}'$ defined by
replacing the kinematical scalar product $(\cdot|\cdot)_{\rm kin}$ with
$(\cdot|\widehat{v^{-1}}\cdot)_{\rm kin}$.
There is a unitary transformation $\hkin'\rightarrow\hkin$ which maps
(\ref{C2}) into (\ref{C1}),  preserves the form of the volume operators
and modifies the holonomy operators appropriately.

\noindent{\bf Acknowledgments} We thank Abhay Ashtekar, Tomasz Paw{\l}owski,
Carlo Rovelli and Thomas Thiemann for discussions about the LQC and the
partial observables respectively. The work was partially supported by the
Polish Ministerstwo Nauki i Szkolnictwa Wyzszego grant  1 P03B 075 29
and  by 2007-2010 research project N202 n081 32/1844.


\begin{thebibliography}{99}
\bibitem{LQGrevsbooks} Thiemann T, Introduction to Modern Canonical Quantum General Relativity (Cambridge: Cambridge University Press) \\
Rovelli C 2004, Quantum Gravity, (Cambridge: Cambridge University
Press) \\ Ashtekar A and Lewandowski J 2004, Background independent
quantum gravity: A status report," {\it Class. Quant. Grav.} \textbf{21}  R53,
(\emph{Preprint} gr-qc/0404018)
%
\bibitem{QG} Rovelli C, Smolin C 1995 Discreteness of area and volume in quantum
gravity, {\it Nucl. Phys.} \textbf{B 442} 593 [1995 Erratum-ibid. \textbf{B
456}  753], (\emph{Preprint} gr-qc/9411005)\\
Ashtekar A, Lewandowski J 1995 Differential Geometry on the
Space of Connections via Graphs and Projective Limits {\it J. Geom.
Phys.} {\bf 17} 191-230, (\emph{Preprint} hep-th/9412073)\\
Ashtekar A and Lewandowski J 1997  "Quantum theory of geometry. I:
Area operators,"{\it  Class. Quant. Grav.} {\bf 14} A55, (\emph{Preprint}
gr-qc/9602046) \\ Ashtekar A and Lewandowski J 1998"Quantum theory of
geometry. II: Volume operators,"{\it  Adv. Theor. Math. Phys.} \textbf{1}
388, (\emph{Preprint} gr-qc/9711031) \\ Ashtekar A, Corichi A and
Zapata J.A 1998 "Quantum theory of geometry. III: Non-commutativity of
Riemannian structures"{\it Class. Quant. Grav.} \textbf{15}  2955,
(\emph{Preprint} gr-qc/9806041)\\ Thiemann T 1998 A length operator
for canonical quantum gravity {\it J. Math. Phys.} \textbf{39}  3372,
(\emph{Preprint} gr-qc/9606092)
%
\bibitem{Qgprop} Thiemann T, 1998 Closed formula for the matrix elements
of the volume operator in canonical
quantum gravity {\it J. Math. Phys.} \textbf{39} 3347, (\emph{Preprint}
gr-qc/9606091)\\
Loll R 1997  Simplifying the spectral analysis of the volume operator
 {\it Nucl. Phys.} \textbf{B 500} 405, (\emph{Preprint} gr-qc/9706038)\\
Brunnemann J and Thiemann T 2006 Simplification of the spectral analysis
of the volume operator in
loop quantum gravity {\it Class. Quant. Grav.} 23  1289,
(\emph{Preprint} gr-qc/0405060)\\ Brunnemann J and Rideout D 2006 Spectral
analysis of the volume operator in loop quantum gravity,
(\emph{Preprint} gr-qc/0612147)\\ Brunnemann J and Rideout D 2007 Properties of
the Volume Operator in Loop Quantum Gravity I: Results,
(\emph{Preprint} 0706.0469)\\ Brunnemann J and Rideout D 2007 Properties
of the Volume Operator in Loop Quantum Gravity II: Detailed
Presentation, (\emph{Preprint} gr-qc/0706.0382.11)
%
\bibitem{QSD} Thiemann T 1998 Quantum spin dynamics (QSD) {\it
Class. Quant. Grav.} \textbf{15} 839, (\emph{Preprint} gr-qc/9606089)\\
Thiemann T 1998 QSD III: Quantum constraint algebra and physical scalar
product in quantum general relativity {\it Class. Quant. Grav.} \textbf{15}
1207, (\emph{Preprint} gr-qc/9705017)
%
\bibitem{DTnewest} Dittrich B and Thiemann T 2007 Are the spectra of geometrical operators in Loop Quantum
Gravity really discrete?, (\emph{Preprint} gr-qc/07081721)
%
\bibitem{ParObs} Rovelli C 2002
Partial observables {\it Phys. Rev. D} \textbf{65} 124013,
(\emph{Preprint} gr-qc/0110035)\\
Dittrich B 2004 Partial and complete observables for Hamiltonian
constrained systems, to
appear in {\it Gen. Rel. Grav. } (\emph{Preprint} gr-qc/0411013)\\
Dittrich B 2006 Partial and Complete Observables for Canonical General
Relativity, {\it Class.
Quant. Grav.} \textbf{23}  6155, (\emph{Preprint} gr-qc/0507106)\\
Thiemann T, 2006 Reduced phase space quantization and Dirac observables
{\it Class. Quant. Grav.} \textbf{23}  1163, (\emph{Preprint} gr-qc/0411031)
%
\bibitem{Rovnewest} Rovelli C 2007 Comment on "Are the
spectra of geometrical operators in Loop Quantum Gravity really
discrete?" by B. Dittrich and T. Thiemann, (\emph{Preprint}
gr-qc/0708.2481)
%
\bibitem{Abhaynewest} Ashtekar A 2007 An Introduction to Loop
Quantum Gravity Through Cosmology, (\emph{Preprint} gr-qc/0702030)
%
\bibitem{LQC}  Bojowald M 2002 Isotropic loop quantum cosmology
{\it Class. Quantum Grav.} \textbf{19}
2717-2741\\
Bojowald M 2005 Loop quantum cosmology {\it Liv. Rev. Rel.} \textbf{8},
11, (\emph{Preprint}
gr-qc/0601085)\\
Ashtekar A, Bojowald M and Lewandowski J 2003 Mathematical structure
of loop quantum cosmology \emph{Adv. Theo. Math. Phys.} \textbf{7}, 233-268, (\emph{Preprint} gr-qc/0304074)
\\ Ashtekar A, Paw{\l}owski T and Singh P 2006, Quantum
nature of the bing bang: An analytical and numerical investigation,
\emph{Phys. Rev.} D\textbf{73} 124038, (\emph{Preprint}
gr-qc/0604013) \\ Ashtekar A, Paw{\l}owski T, and Singh P 2006,
Quantum nature of the big bang, \emph{Phys. Rev. Lett.} \textbf{96},
141301, (\emph{Preprint} gr-qc/0602086) \\ Ashtekar A, Paw{\l}owski
T, Vandersloot K and Singh P 2007, Loop quantum cosmology of $k=1$
FRW models, \emph{Phys. Rev.} D \textbf{75} 024035, \emph{Preprint}
gr-qc/0612104 \\ Szulc {\L}, Kami\'nski W and Lewandowski J 2007,
Closed FRW model in Loop Quantum Cosmology, \emph{Class. Quantum
Grav.} \textbf{24} 2621-2635
\bibitem{APS} Ashtekar A, Paw{\l}owski T and Singh P 2006, Quantum Nature
of Big Bang: Improved dynamics, \emph{Phys. Rev.} D \textbf{74}
084003, (\emph{Preprint} gr-qc/0607039)
%
\bibitem{Thiemmaster} Thiemann T 2006 The Phoenix Project: Master Constraint Programme for Loop Quantum Gravity \emph{  Class.Quant.Grav.} 23  2211-2248
%
\bibitem{KL} Kami{\'n}ski W and Lewandowski J 2007 The flat FRW model in
LQC: the self-adjointness, (\emph{Preprint} gr-qc/0709.3120) \\
Kaminski W, Lewandowski J and Szulc {\L} 2006, Properties of the scalar constraint operator in LQC models, in preparation
%
\bibitem{RovQM} Rovelli C 1991 Quantum Mechanics without Time: A Model
{\it Phys. Rev. D} \textbf{42}  2638\\
Rovelli C 1991 Time in Quantum Mechanics: Physics beyond the Schroedinger
Regime {\it Phys.
Rev. D} \textbf{43} 442.

\end{thebibliography}
\end{document}